\documentclass[aps,pra,showpacs,amsmath,amssymb,floatfix,superscriptaddress,longbibliography,twocolumn]{revtex4-2}
\usepackage{physics}
\usepackage{graphicx}
\usepackage[usenames,dvipsnames,svgnames,table]{xcolor}
\usepackage{tcolorbox}
\usepackage[unicode=true,
            pdfusetitle,
            bookmarks=true,
            bookmarksnumbered=false,
            bookmarksopen=false,
            breaklinks=true,
            pdfborder={0 0 0},
            backref=false,
            colorlinks=true,
            hypertexnames=false]{hyperref}
\hypersetup{linkcolor=NavyBlue,urlcolor=NavyBlue,citecolor=NavyBlue}
\usepackage{microtype}

\begin{document}
\title{Approaching quantum limits in sub-Rayleigh resolution using phase modulation optimization}
\author{Chao-Ning Hu}
\author{Jun Xin}
\email{jxin@hdu.edu.cn}
\author{Xiao-Ming Lu}
\email{lxm@hdu.edu.cn}
\affiliation{School of Physics and Zhejiang Key Laboratory of Quantum State Control and Optical Field Manipulation, Hangzhou Dianzi University, Hangzhou 310018, China}

\begin{abstract}
  Spatial-mode demultiplexing (SPADE) has made significant progress in incoherent optical imaging and has been implemented through various experimental approaches.
  Holography-based SPADE is widely used in proof-of-principle experiments because of its low cost, stable optical arrangement, and ease of implementation.
  However, its low photon utilization severely limits its practical applications.
  Alternating Fourier transforms and phase modulations can implement unitary optical transformations.
  Building on this idea, we use backpropagation to optimize the phase profiles of spatial light modulators (SLMs) and thereby design quantum measurements that are efficient in both information and energy.
  We study centroid--separation estimation for two incoherent point sources, using Fisher information or the estimation errors of an auxiliary neural network estimator as optimization objectives.
  As the number of SLMs increases, the optimized measurements approach the quantum limits characterized by the quantum Cram\'er-Rao bound and the information-regret tradeoff relation.
  With a few SLMs, our method can outperform direct imaging, showing its potential for superresolution imaging.
\end{abstract}

\maketitle
\section{Introduction}
The resolution of conventional optical imaging has long been constrained by diffraction, as characterized by the Rayleigh criterion~\cite{Rayleigh1879}.
However, quantum theory has shown that conventional direct imaging does not necessarily extract all the information available in an optical field.
This insight led to the development of spatial-mode demultiplexing (SPADE)~\cite{Tsang2016,Tsang2017,Tsang2019,Lvovsky2026} to overcome conventional resolution limits.
By measuring suitable spatial modes, SPADE can extract information that is poorly captured by direct imaging.
SPADE has also been implemented and studied in numerous experiments~\cite{Paur2016,Yang2016,Tang2016,Tham2017,Paur2018,Zhou2019,Boucher2020,Tan2023,Santamaria2023,Zhou2023,Rouviere2024,Hu2025,Santamaria2024,Wallis2025a,Kim2025,Liu2026,Paneru2026,Wallis2026}.
Among the available implementations, holography-based SPADE is attractive for proof-of-principle experiments because of its relatively low cost, stable optical arrangement, and ease of implementation.
However, diffraction-order selection and output filtering can discard a substantial fraction of the incident photons~\cite{Arrizon2007,Rosales-Guzman2017}.
Low photon utilization is the Achilles' heel of holography-based SPADE, limiting its usefulness when the available photon budget is small.
This raises the question of whether a photon-efficient measurement can extract all the available information about the parameters of interest, even when several parameters must be estimated together.

To retain the incident photons, we consider unitary optical transformations.
Previous studies have shown that, in an ideal finite-dimensional mode space, a sufficiently long sequence of alternating Fourier transforms and phase modulations can implement an arbitrary unitary transformation~\cite{Morizur2010,LopezPastor2021}.
This idea underlies multi-plane light conversion and its applications in spatial-mode manipulation~\cite{Labroille2014,Fontaine2019,Zhang2023,Kupianskyi2023}.
In principle, such a system, followed by spatially resolved detection, allows us to design quantum measurements that extract information about the parameters of interest~\cite{Warke2026,Filipovich2026}.
The unitary transformation is lossless in principle, but losses from optical components accumulate as more planes are added.
This motivates us to seek phase profiles that yield informative measurements with only a few planes.

Diffractive deep neural networks have demonstrated how multiple optical layers can be trained to perform specific tasks~\cite{Lin2018}.
The layers are optimized jointly by backpropagation through an optical propagation model.
Here, we adopt this architecture and training framework to optimize the phase profiles of spatial light modulators (SLMs) for parameter estimation.
For simplicity, we consider two equally bright incoherent point sources with a Gaussian point-spread function (PSF) and study both separation estimation and joint centroid--separation estimation.
For these tasks, we optimize the measurements using either classical Fisher information or the estimation errors of an auxiliary neural network estimator.
Our numerical results show that even a few modulation planes can outperform direct imaging.
Increasing the number of SLMs allows the separation information to approach its quantum limit at small separations.
For joint estimation, the measurements guided by the auxiliary estimator approach the boundary of the information-regret tradeoff relation (IRTR)~\cite{Lu2021,Hu2025b,Shao2022,Shi2023}.
Meanwhile, the optimized measurements retain nearly all the incident optical energy in the ideal propagation model.

\section{Model and methods}
\subsection{Quantum parameter estimation}
We use two equally bright incoherent point sources to evaluate the optimized optical measurements.
Their image-plane positions are $c-d/2$ and $c+d/2$, where $c$ and $d$ denote the centroid and separation, respectively.
We assume a shift-invariant imaging system with unit magnification and a Gaussian PSF
\begin{equation}
  \psi(x)=\frac{1}{\qty(2\pi\sigma^2)^{1/4}}\exp\qty(-\frac{x^2}{4\sigma^2}),
\end{equation}
where $\sigma$ is the characteristic width.
For weak sources, we describe the spatial information using the state conditioned on the presence of one photon~\cite{Tsang2016}:
\begin{equation}
  \rho_{\theta}\approx\frac12\qty(\op{\psi_{c-d/2}}+\op{\psi_{c+d/2}}),
\end{equation}
where $\ket{\psi_s}=\int\dd{x}\psi(x-s)\ket{x}$ is the single-photon state associated with a point source at position $s$, $\ket{x}$ is the image-plane position basis, and the parameters of interest form the vector $\theta=(c,d)$.
For an unbiased estimator $\hat\theta$, the classical and quantum Cram\'er-Rao bounds give $\operatorname{Cov}(\hat\theta)\succeq F^{-1}\succeq\mathcal F^{-1}$~\cite{Cramer1963,Helstrom1976,Braunstein1994,Kay1993,Liu2019}.
Here $F$ is the classical Fisher information matrix of the chosen measurement, and $\mathcal F$ is the quantum Fisher information matrix of the optical state.
The matrix inequality $A\succeq B$ means that $A-B$ is positive semidefinite.
For the Gaussian PSF, the quantum Fisher information matrix is diagonal~\cite{Tsang2016}:
\begin{equation}
  \mathcal F_{cc} \approx \frac{1}{\sigma^2}\qty(1-\tilde d^2 e^{-\tilde d^2}),\quad \mathcal F_{dd} \approx \frac{1}{4\sigma^2},
\end{equation}
where $\tilde d=d/(2\sigma)$.
Intuitively, classical Fisher information quantifies the information extracted by a quantum measurement, independently of the estimator, while quantum Fisher information gives the information available in the quantum state.

\begin{figure}
  \centering
  \includegraphics{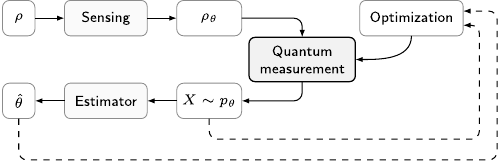}
  \caption{Quantum parameter estimation process and quantum measurement optimization.}
\end{figure}

For single-parameter problems, the quantum Cram\'er-Rao bound is tight.
In the multiparameter case, however, the optimal measurements for different parameters can be incompatible.
To assess this tradeoff, we use the information-regret tradeoff relation (IRTR)~\cite{Lu2021,Hu2025b,Shao2022,Shi2023}.
The normalized-square-root information regrets are defined as
\begin{equation}
  \Delta_c=\sqrt{1-\frac{F_{cc}}{\mathcal F_{cc}}},\quad \Delta_d=\sqrt{1-\frac{F_{dd}}{\mathcal F_{dd}}}.
\end{equation}
Each regret quantifies the information lost relative to the corresponding quantum limit, with zero regret indicating that all the quantum Fisher information for that parameter is extracted.
The IRTR reads
\begin{equation}
  \Delta_c^2+\Delta_d^2+2\sqrt{1-\zeta^2}\Delta_c\Delta_d\geq\zeta^2,
\end{equation}
where $\zeta$ quantifies measurement incompatibility for the two parameters.
For the Gaussian model~\cite{Shao2022},
\begin{equation}
  \zeta^2=\frac{\qty(1-\tilde d^2)^2}{e^{\tilde d^2}-\tilde d^2}.
\end{equation}
For closely spaced sources, the incompatibility coefficient is generally nonzero, indicating that the optimal measurements for centroid and separation estimation are incompatible.
However, at the special separation $d=2\sigma$, $\zeta=0$, so the IRTR permits zero regret for both parameters.
This special case has been studied in detail in Ref.~\cite{Shi2023}.
It is worth noting that the IRTR is not guaranteed to be tight for mixed states~\cite{Lu2021}, but it still provides a useful bound for assessing the information tradeoff.

\subsection{Phase modulation optimization}
\begin{figure}
  \centering
  \includegraphics{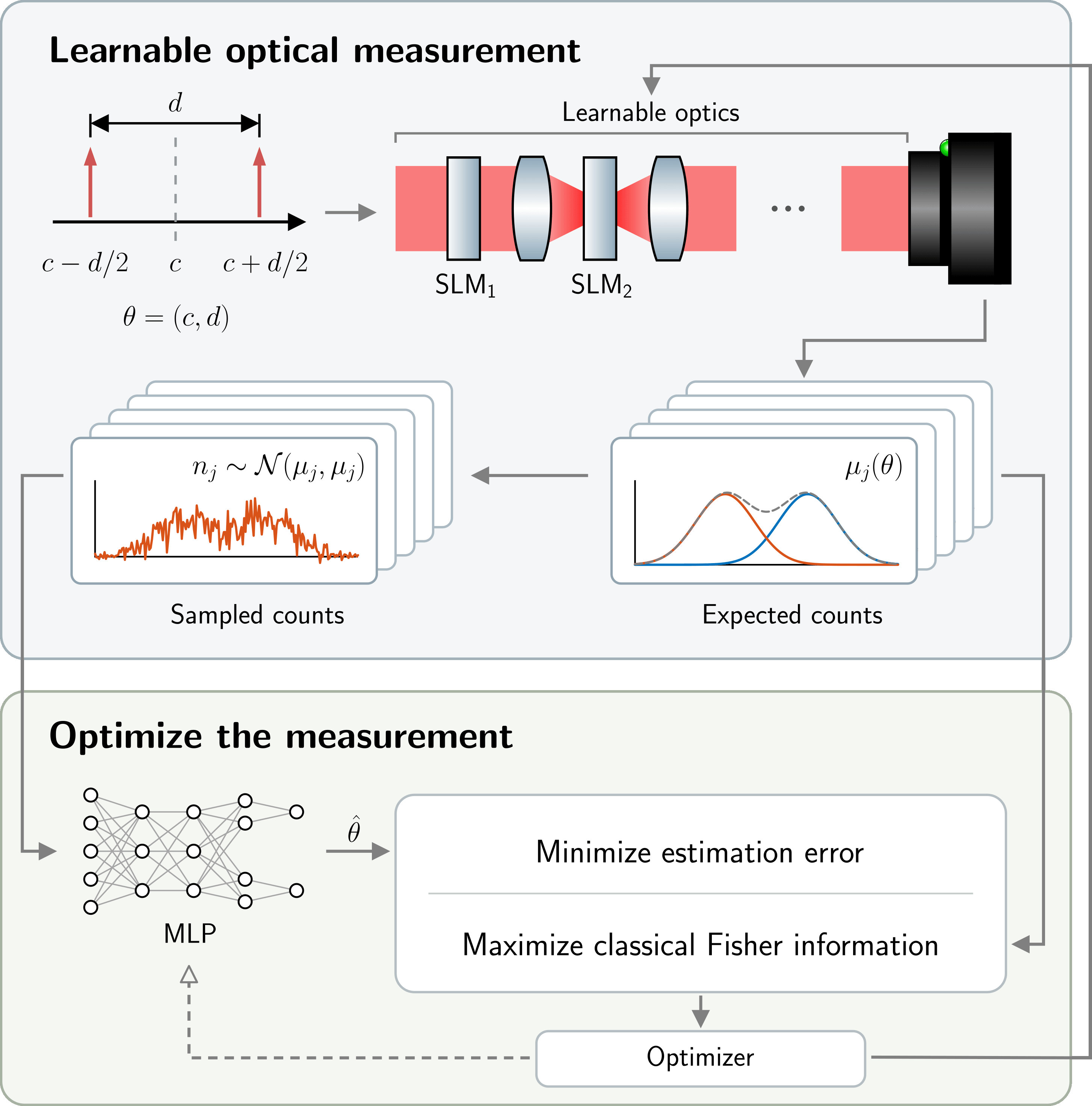}
  \caption{
    Phase modulation optimization.
    The phase profiles of the SLMs are optimized either directly using classical Fisher information or jointly with an auxiliary multilayer perceptron (MLP) using estimation errors.
    Both objectives include a penalty for photon loss.
    Gradients are backpropagated through the optical model to update the phase profiles and, when present, the MLP parameters.
    }
  \label{fig2}
\end{figure}

We design the optical measurement by phase modulation optimization, as illustrated in Fig.~\ref{fig2}.
The system consists of alternating phase-only SLMs and lenses that implement Fourier transforms~\cite{Goodman2017}, followed by a pixelated detector.
The phase profiles determine the measurement and are optimized for the chosen estimation task.

The $\ell$th SLM multiplies the incident field by $e^{i\mathcal G_\ell(x)}$, where $\mathcal G_\ell(x)$ represents the hologram on that SLM~\cite{Rosales-Guzman2017}.
Liquid crystal on silicon (LCoS) SLMs may not faithfully reproduce rapid phase variations because of coupling between liquid-crystal molecules~\cite{Rosales-Guzman2017}.
To favor smoothly varying holograms, we parameterize $\mathcal G_\ell(x)$ using discrete prolate spheroidal sequences (DPSSs), also known as Slepian sequences~\cite{Slepian1978,SciPyDPSS}.
These sequences are orthogonal on a finite sampling grid and are ordered by spectral concentration within a prescribed bandwidth.
We therefore use the leading sequences to help suppress high spatial-frequency components.
The hologram is defined as
\begin{equation}
  \mathcal G_\ell(x)=\pi\tanh\qty[\sum_k a_{\ell k}\phi_k(x)],
\end{equation}
where $\phi_k$ are the DPSS basis functions and $a_{\ell k}$ are the trainable coefficients.
The hyperbolic tangent bounds the phase between $-\pi$ and $\pi$.
Let $\mathcal U$ denote the Fourier transform and $\mathcal M_\ell$ the phase modulation of the $\ell$th SLM.
The total optical propagation is
\begin{equation}
  U=\underbrace{\mathcal U\mathcal M_L\cdots\mathcal U\mathcal M_2\mathcal U\mathcal M_1}_{L~\text{SLMs}}.
\end{equation}
The probability of detecting a photon in the pixel centered at $x_j$ is $p(x_j|\theta)=\tr(\rho_\theta U^\dagger\Pi_j U)$, where $\Pi_j=\int_{D_j}\dd{x}\op{x}$ and $D_j$ is the region covered by that pixel.
The classical Fisher information matrix is calculated from these probabilities.

Phase modulation and Fourier propagation are unitary and photon-number-preserving in the ideal model, so they conserve optical energy.
However, the finite detection window can reduce the collected light.
We penalize this energy loss using $\mathcal L_E=\operatorname{Mean}\qty(1-E_\text{out}/E_\text{in})$, where $\operatorname{Mean}(\bullet)$ denotes the sample mean over a training batch and $E_\text{in}$ and $E_\text{out}$ are the incident and collected optical energies.
The propagation model does not include optical losses in the SLMs themselves.

We consider two ways to optimize the measurement. 
The first uses classical Fisher information directly.
For separation estimation with known centroid $c=0$, let $F_d$ be the classical Fisher information about $d$ and $\mathcal F_d=1/(4\sigma^2)$ its quantum limit.
Let us denote the normalized information regret by $\tilde R_d=1-F_d/\mathcal F_d$ to quantify the information loss. 
We minimize
\begin{equation}
  \mathcal L_F=\operatorname{Mean}\qty[w(d)\tilde R_d]+\mathcal L_E,
\end{equation}
where $w(d)=e^{-\eta d/\sigma}$ with $\eta=2$ gives greater weight to smaller separations.

The second approach optimizes the measurement jointly with an auxiliary multilayer perceptron (MLP) estimator.
An MLP provides fast, differentiable estimates as the measurement changes during training, avoiding the repeated numerical optimization required by maximum likelihood estimation (MLE).
A more expressive estimator could reduce the training error without improving the optical measurement.
We therefore use a compact network to limit model capacity.
The hidden layers are also much narrower than the detector input, creating a bottleneck intended to reduce reliance on prior information from the training distribution.
Together, these designs help focus the optimization on the optical measurement rather than on the estimator alone.
The MLP architecture and numerical settings are given in Sec.~\ref{sec:results}.

For training, we approximate the photon counts as $n_j\sim\mathcal N(\mu_j,\mu_j)$, where $\mu_j=Np(x_j|\theta)$ and $N$ is the mean incident photon number.
This approximation reproduces the mean and variance of Poisson counts.
It also allows direct backpropagation through sampling without a straight-through estimator or an additional gradient approximation.
The MLP maps the sampled photon counts to the estimates.

For single-parameter separation estimation with the centroid known and fixed at $c=0$, the MLP outputs $\hat d$, and the loss is
\begin{equation}
  \mathcal L_d=\tilde{\mathcal E}_d+\mathcal L_E,
\end{equation}
where $\tilde{\mathcal E}_d$ is the weighted mean-squared error (MSE) of $\hat d$, averaged over the training batch and normalized by $\sigma^2$.
The weight $w(d)$ favors estimation at small separations.
For multiparameter estimation of $c$ and $d$, the MLP outputs both $\hat c$ and $\hat d$.
We minimize
\begin{equation}
  \mathcal L_{cd}=\log\qty(\mathcal E_c^{1-\lambda}\mathcal E_d^\lambda) + \mathcal L_E,
\end{equation}
where $\mathcal E_c$ and $\mathcal E_d$ denote the MSE of $\hat c$ and $\hat d$, respectively, normalized by $\sigma^2$ and computed with uniform weights over the training samples.
Here $\lambda$ is a hyperparameter that controls the relative weights of the two errors, and we set $\lambda=0.9$.
We use the geometric mean to reduce the influence of absolute error scales and take its logarithm for computational convenience.

We use backpropagation to optimize the phase coefficients and, when present, the MLP parameters.
Note that the auxiliary MLP is not guaranteed to be unbiased and may use prior information from the training distribution.
It serves only to guide measurement optimization.
After training, we evaluate the optimized measurement using classical Fisher information and information regrets, which are calculated from the detection probabilities and do not depend on the estimator.
In principle, the auxiliary MLP can be discarded after training, and parameter estimates can instead be obtained by MLE or a separately trained neural network for the fixed measurement.

\section{Numerical results}\label{sec:results}
We use a wavelength of $770\,\text{nm}$, a lens focal length of $150\,\text{mm}$, and a Gaussian PSF with $\sigma=100\,\text{\textmu m}$.
The mean incident photon number is $N=1000$.
Each optical plane is sampled at 512 points with a spacing of $4.6\,\text{\textmu m}$, covering a range of approximately $\pm11.8\sigma$.
We use a standardized half-bandwidth of 235.52 and retain the first 471 DPSSs~\cite{SciPyDPSS}.
The coefficients $a_{\ell k}$ are initialized independently from $\mathcal N(0,0.1^2)$.

Direct imaging serves as the reference measurement.
For separation estimation, we fix the centroid at $c=0$ and sample the training separations uniformly from $0<d<3\sigma$.
For multiparameter estimation, we optimize a separate measurement at each working point.
The training values of $c$ and $d$ are sampled uniformly within $\pm0.1\sigma$ of the corresponding true values.

The MLP has 512 inputs and 2 hidden layers of 16 neurons each, shared by 2 output heads for multiparameter estimation.
Each head contains 8 hidden neurons and a linear scalar output.
All hidden layers use rectified linear unit (ReLU) activations.
We use the Adam optimizer with a learning rate of $10^{-3}$ on an RTX 4090 GPU.
Each optimization uses $2^{20}$ training parameter samples and a batch size of $2^{10}$ over 32 epochs (8 for $\mathcal L_d$).

\subsection{Separation estimation}
\begin{figure}[b]
  \centering
  \includegraphics{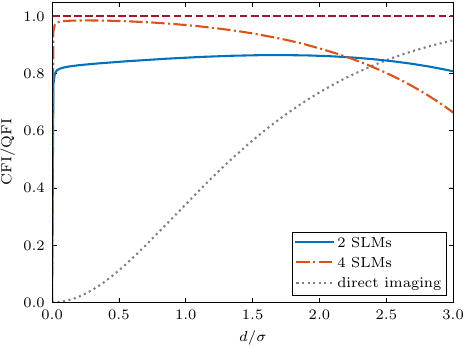}
  \caption{
    Fisher information about the separation $d$ for direct imaging and measurements optimized with 2 and 4 SLMs by minimizing $\mathcal L_F$.
    CFI and QFI denote classical and quantum Fisher information, respectively.
    }
  \label{fig3}
\end{figure}

Figure~\ref{fig3} compares the classical Fisher information about $d$ for direct imaging and measurements optimized using $\mathcal L_F$.
At $d\approx0.1\sigma$, measurements with 2 and 4 SLMs retain approximately $82\%$ and $98\%$ of the quantum Fisher information, respectively, compared with less than $1\%$ for direct imaging.
With 4 SLMs, the information remains close to the quantum limit over a range of small separations.
Similar results were reported in Ref.~\cite{Filipovich2026}.
In that work, the classical Fisher information was averaged uniformly over separations, whereas we use $w(d)$ to emphasize small separations.

\begin{figure}[h]
  \centering
  \includegraphics{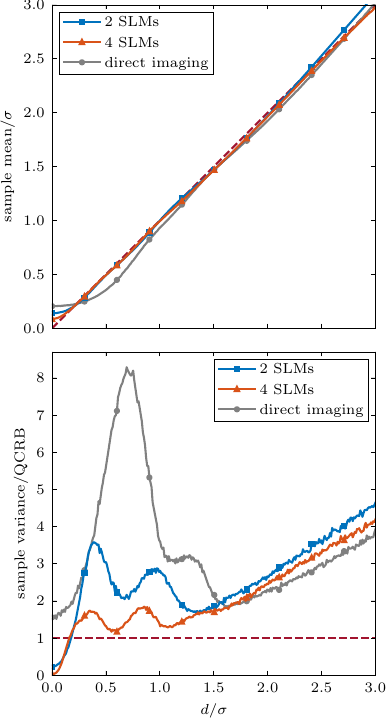}
  \caption{
    Sample mean (top) and rescaled sample variance (bottom) of the MLP separation estimates $\hat d$ for direct imaging and measurements optimized with 2 and 4 SLMs using $\mathcal L_d$.
    Each separation is evaluated with $10^4$ independent simulated photon-count records generated using the Gaussian sampling approximation.
    The sample variance is multiplied by the mean photon number $N$ and normalized by the quantum Cram\'er-Rao bound (QCRB).
    }
  \label{fig4}
\end{figure}

Figure~\ref{fig4} shows the sample mean and variance of the separation estimates obtained by minimizing $\mathcal L_d$.
For each measurement, including direct imaging, we train a separate MLP using the same architecture, loss, and training distribution.
For direct imaging, the MLP is used without phase modulation.
At small separations, optimized phase modulation combined with an MLP gives smaller bias and variance than the MLP alone.
For example, at $d\approx0.5\sigma$, the rescaled sample variance with 4 SLMs is about $1.3$ times the quantum Cram\'er-Rao bound, compared with about $5.7$ times for the MLP trained on direct imaging records.
The difference becomes smaller at larger separations.
At very small separations, the rescaled sample variance falls below the quantum Cram\'er-Rao bound.
This is because the MLP estimates are biased, whereas the bound shown in the figure applies to unbiased estimators.

\subsection{Centroid and separation estimation}

\begin{figure*}
  \centering
  \includegraphics{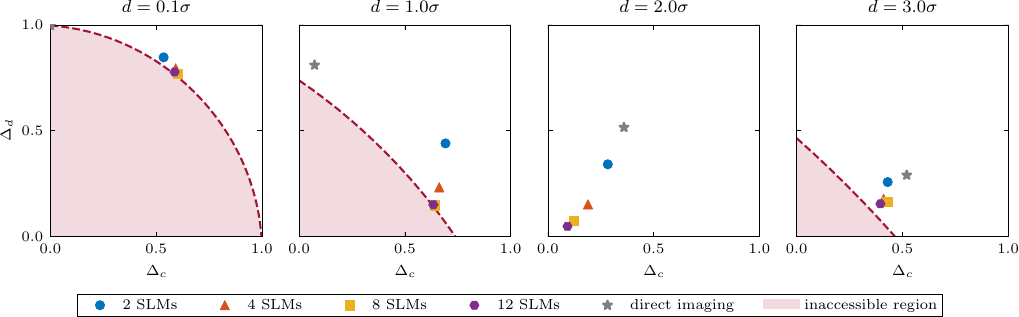}
  \caption{
    Normalized-square-root information regrets for centroid and separation estimation at $d/\sigma=0.1,1,2,3$, with $c=0$.
    The phase profiles are optimized separately for each working point.
    Dashed curves show the IRTR boundary, and shaded regions are excluded by the relation.
    }
  \label{fig5}
\end{figure*}

Figure~\ref{fig5} compares the information regrets of the optimized measurements with those of direct imaging for joint estimation of $c$ and $d$.
For each working point, we train a separate measurement by minimizing $\mathcal L_{cd}$ over a small neighborhood of that point, then fix the phase profiles and calculate the classical Fisher information matrix and information regrets at the working point itself.
At $d=0.1\sigma$ and $d=\sigma$, the optimized measurements extract more separation information than direct imaging, but this gain comes at the cost of less centroid information.
In the figure, this corresponds to a smaller separation regret and a larger centroid regret.
With 8 and 12 SLMs, the optimized measurements approach the IRTR boundary at both separations.
These results show that the estimation errors of an auxiliary MLP can guide phase optimization toward measurements close to the quantum limit on this information tradeoff.
  
At $d=2\sigma$, the incompatibility vanishes, and both parameters can benefit from the optimized measurement.
As the number of SLMs increases from 2 to 12, the regrets decrease from approximately $(\Delta_c,\Delta_d)=(0.28,0.34)$ to $(0.092,0.048)$.
The measurement with 12 SLMs extracts about $99.1\%$ and $99.8\%$ of the quantum Fisher information for the centroid and separation, respectively.
At $d=3\sigma$, all optimized measurements also extract more information about both parameters than direct imaging.

The optimized measurements also retain high photon utilization.
In this model, more than $99\%$ of the incident energy is collected for all evaluated configurations, both on the single-parameter evaluation grids and at the joint-estimation working points.
The information gains therefore come with little photon loss in the ideal propagation model.

\section{Conclusion}
In this work, using two incoherent point sources as an example, we use backpropagation to optimize the SLM phase profiles in a sequence of alternating Fourier transforms and phase modulations. 
As the number of SLMs increases, the optimized measurements approach the quantum limits set by the quantum Cram\'er-Rao bound and the IRTR, while retaining nearly all incident optical energy in the ideal propagation model. 
We consider two objectives: maximizing the classical Fisher information directly, and using the estimation errors of an auxiliary neural-network estimator to guide measurement optimization.
Both outperform direct imaging in the tasks considered.

Although we illustrate the approach with a simple model of two incoherent point sources, the method can in principle be extended to other parameters, such as the length of a line source or the width of a two-dimensional source, and to problems with more unknown parameters.
The auxiliary-estimator approach still relies on supervised learning, and developing unsupervised or self-supervised schemes that exploit information from the physical process remains an open question. 
In addition, the phase profiles are optimized entirely in idealized simulations, so the resulting measurements may be sensitive to noise, misalignment, and other experimental imperfections.
Accounting for these effects and realizing the optimized measurements on a real optical system remains to be explored in future work.

\section*{Acknowledgments}
This work is supported by the Quantum Science and Technology-National Science and Technology Major Project (Grant No. 2024ZD0301000), the National Natural Science Foundation of China (Grants No. 92476118 and No. 12275062), and Key R\&D Program of Zhejiang (Grant No. 2026C01004).

\section*{Data availability}
The data in this study were generated by numerical simulations.
The source code is available in our GitHub repository~\cite{Code}.

\end{document}